# A defect-free naphthalene diimide bithiazole copolymer via regio-selective direct arylation polycondensation

Rukiya Matsidik[a], Michele Giorgio [b,c], Alessandro Luzio[b], Mario Caironi[b], Hartmut Komber[d], and Michael Sommer*[a]

[a]     Dr. Rukiya Matsidik, Prof. Michael Sommer
Institute for Macromolecular Chemistry, University of Freiburg, Stefan-Meier Str. 31, 79104 Freiburg, Germany
*E-mail: michael.sommer@chemie.tu-chemnitz.de

[b]     Michele Giorgio, Dr. Alessandro Luzio, Dr. Mario Caironi
Center for Nano Science and Technology @PoliMi, Istituto Italiano di Tecnologia, Via Pascoli 70/3, 20133 Milano, Italy

[c]     Michele Giorgio
Dipartimento di Elettronica, Informazione e Bioingegneria, Politecnico di Milano, P.za L. Da Vinci, 32, 20133 Milano, Italy

[d]     Dr. Hartmut Komber
Leibniz-Institut für Polymerforschung Dresden e.V., Hohe Straße 6, 01069 Dresden, Germany

**Abstract**

An optimized direct arylation polycondensation (DAP) protocol for the synthesis of a novel naphthalene diimide (NDI) 2,2´-bithiazole (2-BTz) copolymer (PNDI-2-BTz) is presented. The regioselective C-H activation of 2-BTz at the 5-positions allows for the synthesis of fully regioregular and homocoupling-free PNDI-2-BTz of high molecular weight in less than 1 h in quantitative yield. Complete end group assignment shows functionalities according to monomer structures or to nucleophilic substitution, and allows for the reliable determination of absolute molecular weight. Compared to the well-known bithiophene analog PNDIT2, an exceptionally high thermal stability, a hypsochromically shifted charge transfer absorption band and a lower-lying LUMO energy level is found, making PNDI-2-BTz an interesting candidate for applications in organic electronic devices. In contrast to the selective and high yielding C-H activation of 2-BTz at the 5-position, the regioisomer 5,5´-bithiazole is inactive under a variety of conditions.

**Introduction**

Naphthalene diimide (NDI) based π-conjugated materials have gained extensive interest for organic electronics owing to their unique properties such as low electron affinity, simple synthesis, ease of structural modification and excellent electrical properties.[1–4] Represented in many cases by the NDI bithiophene (T2) copolymer, hereafter referred to as PNDIT2, NDI copolymers are prime candidates for applications that require high electron mobilities.[1,4,5] The common structural design strategy for these NDI-based n-type materials is the donor acceptor (D-A) approach combining NDI with strong donors, in many cases based on thiophene[1], to create intramolecular charge transfer states along the polymer backbone, leading to widely tunable optical absorption from the Vis up to the NIR region. This aspect has enabled usage of NDI copolymers as n-type materials in high-performance all-polymer photovoltaics.[6,7] While for photovoltaics tunable absorption in the visible region is an important factor, for doping applications stability of the radical anion in ambient conditions is of particular importance. With the position of the LUMO being the decisive parameter for radical anion air stability, a stabilization of the LUMO can be achieved either by derivatizing the NDI or the donor comonomer (the latter being however less effective than the former). Several derivatives of PNDIT2 have been presented in terms of structural variation of the donor comonomer including biselenophene[8] and bifuran[9], but the NDI copolymer with 2,2´-bithiazole (2-BTz) has not yet been presented.

2-BTz is interesting as this comonomer is less electron rich than bithiophene and thus expected to slightly lower the LUMO level of the copolymer at an otherwise retained dihedral NDI/2-BTz torsion angle.[10–12] For 5-BTz an additional NDI/5-BTz planarization is expected due to reduced steric hindrance at the imide oxygens. Such effects are also known from other BTz-based π-conjugated polymers.[10,13,14] Additionally, owing to possible S and N interactions, BTz-based polymers have been reported to show strong interchain interactions that eventually lead to a higher degree of crystallinity.[10,15–17] With these advantageous properties, BTz became an interesting building block for materials with applications in organic field effect transistors (OFETs) and organic photovoltaics (OPVs).[12,13,16,18]

From synthetic point of view, state-of-the-art Stille or Suzuki polycondensations have been mostly applied to prepare BTz-[11,19–21] and also NDI-based[1,4] polymers. These synthetic transformations are currently being replaced by in many ways more attractive C-H functionalization strategies. This approach not only reduces the overall number of synthetic steps and costs but also eliminates the problems associated with other synthetic methods.[22–24] In this regard, both the known low stability of stannylated BTz comonomers[10,25] as well as the sufficiently high C-H reactivities of non-functionalized BTz[10,10,14,26] make this building block most ideal for C-H activation polycondensation, hereafter referred to as direct arylation polycondensation (DAP).[10,14]

Typical protocols for the DAP of 2-BTz employ N,N-dimethyl acetamide (DMAc) as solvent. Under such conditions low molar mass materials are unavoidably obtained upon copolymerization with 2,6-dibromo naphthalene diimide (NDIBr$_2$) due to both low solubility of the resulting copolymers and as NDIBr$_2$ (and –NDIBr chain ends) have a strong tendency to undergo nucleophilic substitution to yield NDI-OH termini.[27–29] By replacing DMAc with aromatic solvents we have recently established a simple and highly efficient DAP protocol for PNDIT2 and other NDI copolymers.[28,30]

Here, an extension towards 2-BTz is presented. In contrast to PNDIT2 synthesis via DAP[28], copolymerization of 2-BTz and NDIBr$_2$ requires tris(*o*-anisyl)phosphine (P(*o*-anisyl)$_3$) as ligand. Under these conditions, regio-isomerically pure and homocoupling-free PNDI-2-BTz is obtained in quantitative yield. The robust protocol allows to adjust monomer concentration such that very short reaction times down to 20 min become possible, which is unusual for a polycondensation process (Scheme 1). Characterization of the thermal, optical and electrical properties suggests this material to be a promising candidate for all-polymer solar cells, field-effect transistors and doping applications.

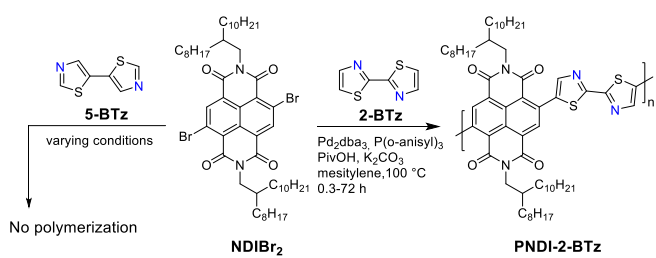

**Scheme 1.** Synthetic route for the DAP of 2-BTz and NDIBr$_2$. For reaction conditions intended to copolymerize NDIBr$_2$ and 5-BTz, see Table S1.

**Results and Discussion**

**Synthesis of polymers**

The C-H reactivity and regioselectivity of heterocycles is significantly influenced by the heteroatom. For thiophene derivatives, S-Pd precoordination is thought to cause high reactivity and selectivity at the α-C-H position.[26] The extraordinarily robust DAP protocol for the synthesis of PNDIT2[28,30] was applied to copolymerize NDIBr$_2$ and 2-BTz. However, no reaction occurred with both monomers remaining intact (Table 1, entry 1). This points to complexation of Pd(0) with 2-BTz, thus prohibiting oxidative addition. Thus, DAP reaction conditions were sought that make use of additional phosphines, such as the Herrmann Beller (HB) catalyst combined with P(*o*-ansiyl)$_3$.[23] Although the reaction mixture displayed a light pink color indicating a donor-acceptor compound, most of the monomer was recovered except some amount of hydroxylated NDI as a result of nucleophilic substitution (Table 1, entry 2).[28] As this reaction can also take place from species in which Pd is oxidatively inserted in the NDI-Br bond,[27] we envisioned an additional phosphine ligand to be useful to displace the potentially coordinating 2-BTz monomer from Pd(0). We thus turned back to Pd$_2$dba$_3$ in combination with P(*o*-ansiyl)$_3$ as ligand (Table 1, entries 3-5). At a concentration [NDIBr$_2$] of 0.1, the reaction mixture gelated after less than 1 h indicating a very fast reaction, and a red solid was obtained in 100 % yield after Soxhlet extraction with a nominally high molecular weight $M_{n,SEC}$ of 89 kg/mol. Further variation of monomer concentration allowed molar mass to be tuned (Table 1, entries 4,5). Polymerization at 0.2 M led to products that were only soluble in hot chlorobenzene, while the DAP at 0.05 M yielded a moderate $M_{n,SEC}$ = 40 kg/mol. In the latter case, the mixture was terminated after 72 hours by cooling, whereas in the other cases the mixture was cooled after gelation followed by work up.

Having achieved an unusually efficient coupling between 2-BTz and NDIBr$_2$ with tunable molecular weight in quantitative yield, we turned to the copolymerization of NDIBr$_2$ and the regioisomeric 5-BTz. However, none of the conditions from Table 1 was successful as 5-BTz remained entirely unreacted. According to the calculated free energy of the C-H bond activation using a model catalytic system, the C-H bond at 2-position of thiazole is significantly less reactive than the one in 5-position (26.3 kcal/mol versus 23.7 kcal/mol, respectively), but still sufficiently reactive when compared to common C-H substrates such thiophene or furan, in which the α-C-H bonds have 25.6 kcal/mol and 25.3 kcal/mol, respectively.[26] Experimental studies reveal a somewhat more complex behavior. The trend of a higher C-H bond reactivity at 5-position is experimentally only observed for DA conditions involving polar solvents and very bulky phosphines, while for non-polar solvents and less bulky ligands such as P(*o*-ansiyl)$_3$ activation of the 2-position is favoured.[31] Based on the study by Chavez et al., the herein reported conditions in mesitylene appeared promising also for the copolymerization of 5-BTz.[31] However, neither the conditions of entries 1-5 nor variations involving co-catalysts such as CuI, Cu(OAc)$_2$ or Ag salts did lead to coupling of NDIBr$_2$ and 5-BTz (Table S1 supporting information).[32,33]

**Table 1.** Screening the reaction parameters for PNDI-2-BTz. All entries were run at 100 °C in mesitylene, and 3.0 eq. of K$_2$CO$_3$ as base, 1.0 eq. of PivOH as additive were used. P(o-ansiyl)$_3$ was used as ligand.

| Entry | Cat./ mol% | Ligand [mol%] | [M][a] | T [h] | Yield [%] | M$_n$/M$_w$[b] [kDa] | DP$_{NMR}$[c] |
|---|---|---|---|---|---|---|---|
| 1 | Pd$_2$dba$_3$/1.0 | 0 | 0.2 | 72 | 0 | 0 | 0 |
| 2 | HB/2.0 | 4.0 | 0.1 | 72 | 0 | 0 | 0 |
| 3 | Pd$_2$dba$_3$/2.0 | 4.0 | 0.1 | 0.8 | 100 | 89/444 | 28 |
| 4 | Pd$_2$dba$_3$/2.0 | 4.0 | 0.05 | 72 | 98 | 40/210 | 24 |
| 5 | Pd$_2$dba$_3$/2.0 | 4.0 | 0.2 | 0.3 | 99 | n.d. | 63 |

[a] Concentration of monomer e.g. NDIBr$_2$ in mesitylene (mol/L); [b] Size exclusion chromatography (SEC) data in CHCl$_3$ at room temperature; [c] Degree of polymerization calculated from $^1$H NMR signal intensities (see SI);

**NMR analysis**

To determine the regioregularity of PNDI-2-BTz and possibly end groups allowing for absolute molecular weight determination, high-temperature NMR analysis was carried out at 120 °C in C$_2$D$_2$Cl$_4$. Narrow signals were found for NDI and 2-BTz backbone protons suggesting molecularly dissolved chains. In addition, only expected signals derived from either 2-BTz or from NDI-OH and occasionally also from NDI-OPiv chain ends were seen, with a total intensity corresponding to the trend of molar masses obtained from SEC (Table 1, Figure 1).[28,29] Such a fortunate situation is again unusual for polycondensation products and allows for the determination of absolute degrees of polymerization (DP$_{n,NMR}$) and number average molecular weights ($M_{n,NMR}$). Integration of entries 3, 4 and 5 (Figure S1-S3) gave DP$_n$ = 28, 24 and 63, corresponding to $M_{n,NMR}$ = 27.8, 23.8, and 62.5 kDa, respectively (Table 1, for integration and calculation see Supporting Information). From the comparison of SEC- and NMR-based molecular weights, the typical overestimation by SEC arising from i) stiff chains and ii) additional aggregation in solution (Figure S7) is seen. The presence of narrow backbone signals in addition to the unequivocal assignment of end groups does not leave room for any defect, and thus proves a fully regioregular, homocoupling-free structure.

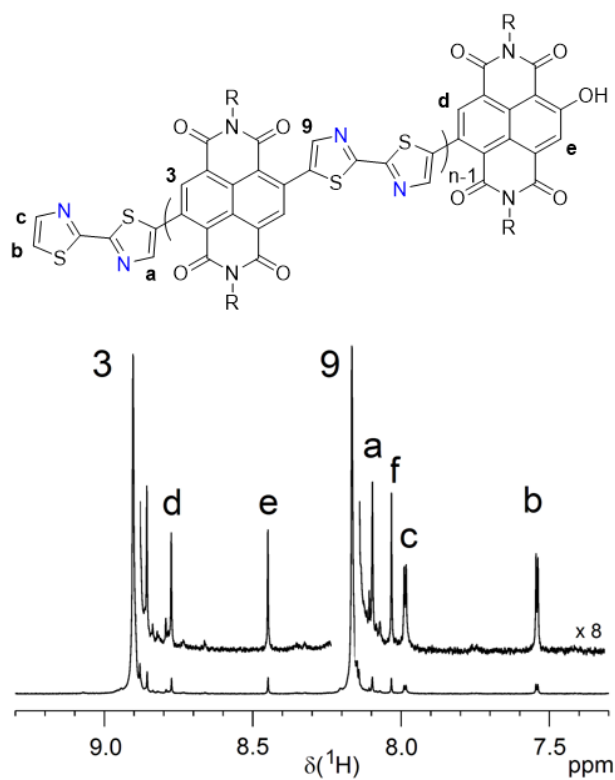

**Figure 1.** $^1$H NMR spectrum (region) of entry **4** (Table 1) showing the backbone signals (3, 9) and end groups signals (a – f). Signal f results from the Tz unit next to the NDI-OH end group. R = 2-octyl dodecyl. Solvent: $C_2D_2Cl_4$ at 120°C.

**Optical and thermal properties**

Optical absorption of PNDI-2-BTz in solution featured a typical donor-acceptor system having high energy π-π* band and low energy charge-transfer (CT) band. As expected, compared to PNDIT2, PNDI-2-BTz showed a hypsochromically shifted CT band due to 2-BTz being a weaker donor than T2 (Figure 2a). The hypsochromic shift of ~100 nm is rather high but not unexpected when e.g. compared to similar bithiophene/ bithiazole copolymers.[20] As many high performance donor polymers show absorption between 550 nm to 850 nm,[34] PNDI-2-BTz would be a promising complementary absorption acceptor material for e.g. PTB7.[35,36] The LUMO energy level of PNDI-2-BTz was extracted from cyclic voltammetry (CV) measurements and estimated as -3.83 eV, which is slightly lower compared to PNDIT2 (-3.78 eV) (Figure 2). In addition to the for NDI materials usual two electron reductions, the CV curve of PNDI-2-BTz exhibits a shoulder of minor intensity at ~ -0.7 V, indicating irreversible processes of unknown nature.

The thermal stability of PNDI-2-BTz was analyzed using thermal gravimetric analysis (TGA) and differential scanning calorimetry (DSC). From TGA a high thermal stability of PNDI-2-BTz with an onset temperature $T_{on}$ ~ 445 °C is found (Figure S8). DSC curves of PNDI-2-BTz are featureless showing no significant thermal transition such as melting or crystallization (Figure S9).

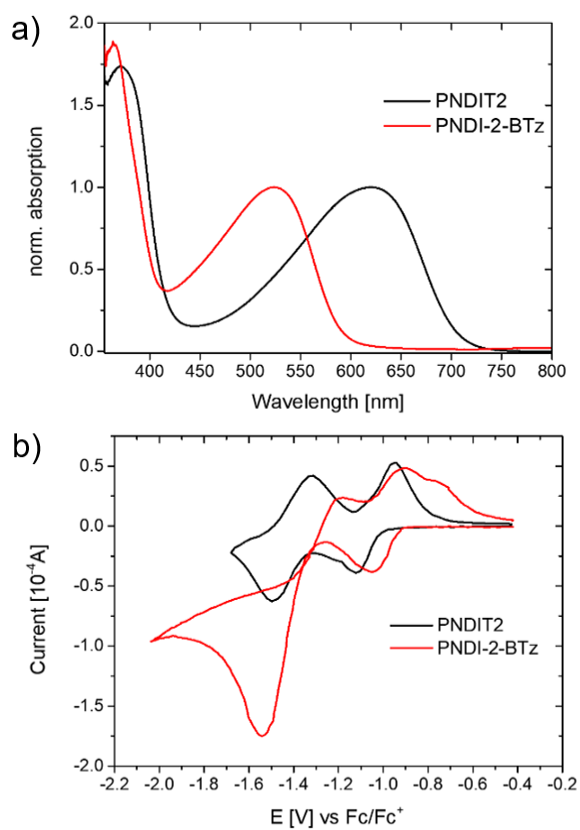

**Figure 2.** a) UV-vis absorption of PNDI-2-BTz and PNDIT2 in 1-chloronaphthalene at room temperature. b) CV curves of PNDIT2 and PNDI-2-BTz films in 0.1 M TBAPF$_6$ acetonitrile solution at a scan rate of 100 mV/s.

**Electron transport properties**

To assess the electron transport properties of PNDI-2-BTz films, we fabricated bottom contact, top-gate field-effect transistors (FETs). Both a common spin-coating procedure, as well as an off-center spin coating method[37] were applied to deposit solutions of PNDI-2-BTz to induce uniaxial chain alignment.[38] The deposition was followed by annealing at different temperatures between 150°C - 350 °C. The transfer characteristics were measured in the saturation regime, with $V_{DS}$ = 60 V (Figure 3a). For all samples, quasi ideal n-channel field effect behavior was observed as evidenced by the typical transfer curve in the electron accumulation regime reported in Figure 3a.

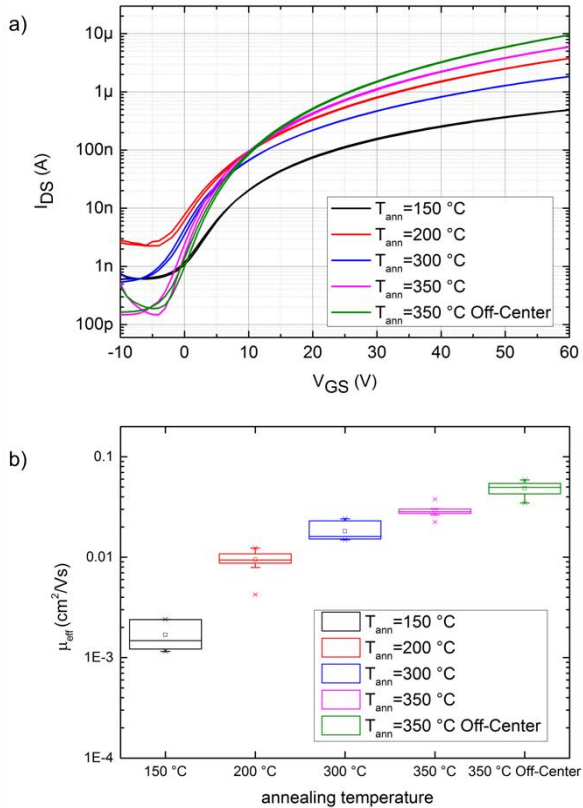

**Figure 3.** a) Typical transfer characteristics of PNDI-2-BTz based FETs. b) Effective saturation mobilities as a function of processing conditions.

It is found that the transport properties of films improve upon annealing up to 350 °C, leading to an increase of the ON currents from 0.5 µA to 5 µA. Transistor ON current is further increased upon alignment parallel to the channel length direction using off-center spin coating, reaching 10 µA.[37] Field effect mobility values, $\mu_{sat}$ were extracted from the slope of the transfer curves according to the gradual channel approximation. Since the square root of the transfer curve in saturation is not always linear over the whole investigated range, we used, as a more reliable parameter, the effective mobility $\mu_{eff}$ as the product between $\mu_{sat}$ and the measurement reliability factor r, as recently suggested by Choi *et al*. (Figure 3b).[39] As a direct consequence of what already observed for transfer curves, $\mu_{eff}$ ranges from less than 3×10$^{-3}$ cm$^2$/Vs, if the film is annealed at 150° C, to 0.03 cm$^2$/Vs, if the film is annealed at 350° C. Mobility is further improved up to 0.05 cm$^2$/Vs, with a maximum mobility of 0.06 cm$^2$/Vs, when the film is deposited using off-center spin coating and annealed at 350 °C. All extracted mobility values are reported in Table 2. It is apparent that electron mobilities are lower compared those previously reported for PNDIT2.[28] Shining light onto this discrepancy will require combined morphological and theoretical investigations.

**Table 2.** $\mu_{sat}$, r and $\mu_{eff}$ extracted from current-voltage curves in Figure 3a.

| Deposition conditions | $\mu_{sat}$ [cm$^2$/Vs] | r | $\mu_{eff}$ [cm$^2$/Vs] |
| --- | --- | --- | --- |
| 150 °C | 0.003 | 1 | 0.003 |
| 200 °C | 0.009 | 0.97 | 0.009 |
| 300 °C | 0.025 | 0.73 | 0.018 |
| 350 °C | 0.045 | 0.67 | 0.030 |
| 350 °C off-center | 0.085 | 0.57 | 0.048 |

**Conclusions**

We have presented an optimized direct arylation polycondensation (DAP) protocol for the synthesis of the novel naphthalene diimide (NDI) 2,2´-bithiazole (2-BTz) copolymer PNDI-2-BTz. Full regioselective C-H activation of 2-BTz at the 5-position allows for the synthesis of regioregular, defect-free copolymers of high and tunable molecular weight in quantitative yield and very short reaction times. NMR end group analysis gives access to absolute molecular weights which, compared to the significantly larger values from size exclusion chromatography, point to strong aggregation in solution at room temperature. The exceptionally high thermal stability, the hypsochromically shifted charge transfer absorption band as well as the low-lying LUMO energy level compared to common NDI copolymers render PNDI-2-BTz an interesting candidate for applications in organic electronic devices such as transistors, thermoelectric devices and solar cells. The regioisomeric PNDI-5-BTz could not be synthesized using a variety of DAP conditions, but can be expected to show similarly interesting properties due to possible coplanarization of NDI-5-BTz linkages.

**Experimental Section**

**Materials and methods**. *Materials*: All chemicals were obtained from Sigma Aldrich and used without further treatment unless otherwise stated. $Pd_2dba_3$ was obtained from Sigma Aldrich in 97 % purity and was used as received, and stored under ambient conditions. 2,2′-Bithiazole was purchased from ABCR (98 %) and further purified by eluting through a silica plug with *iso*-hexanes and diethyl ether (5:1). Monomer $NDIBr_2$ [40] and 5,5′-bithiazole[41] were synthesized according to previously reported methods. *Instrumentation*: SEC measurements were carried out on four SDV gel 5 μm columns, with pore sizes ranging from $10^3$ to $10^6$ Å (PSS), connected in series with a Knauer K-2301 RI detector, and calibrated with polystyrene standards. $CHCl_3$ was used as eluent at room temperature at a flow rate of 1.0 mL/min. UV-vis measurements were carried out on a Perkin Elmer $\lambda_{1050}$ spectrophotometer, using a tungsten lamp as the excitation source. $^1H$ (500.13 MHz) and $^{13}C$ (125.76 MHz) NMR spectra were recorded on a Bruker Avance III spectrometer using a 5 mm gradient probe. DSC measurements were acquired on a NETZSCH DSC 204 F1 Phoenix under a nitrogen atmosphere at a heating and cooling rate of 10 K/min.

**General procedure for polymerization** (Table 1, entry 3). To a dry vial containing a magnetic stir bar 2-BTz (26.9 mg, 0.16 mmol), $NDIBr_2$ (157.6 mg, 0.16 mmol), $K_2CO_3$ (66.3 mg, 0.48 mmol), and pivalic acid (16.3 mg, 0.6 mmol) were carefully added followed by the addition of 1.6 mL degassed mesitylene under a $N_2$ atmosphere, and the whole was stirred for a couple of minutes at RT to fully dissolve the monomers. Then $Pd_2dba_3$ (2.9 mg, 0.0032 mmol) and P(*o*-anisyl)$_3$ (2.3 mg.0064 mmol) were added under nitrogen. The vial was sealed and placed into a preheated oil bath and stirred at 100 °C until the reaction mixture gelated (50 minutes). After cooling to RT, the material was dissolved in 30 mL $CHCl_3$, precipitated into 200 mL methanol, filtered and purified via Soxhlet extraction with acetone, ethyl acetate, and hexanes (until colorlessness of the extraction solution). The material was finally collected with $CHCl_3$ to give 157 mg PNDI-2-BTz in quantitative yield. SEC (25°C, $CHCl_3$): $M_n$ = 89 kDa, $M_w$= 444 kDa.

**OFETs fabrication.** Bottom gold contacts (30 nm) were fabricated on glass substrates using conventional photolithography. Semiconducting polymer was dissolved in chlorobenzene at a concentration of 5 mg/mL, and spin coated at 1000 rpm for 30 s. After deposition the samples were annealed at different temperatures (150°C ~ 350 °C) for 30 min starting from room temperature and gradually increasing the temperature by 10 K/min. Then poly(methylmethacrylate) (PMMA) dissolved in n-butyl acetate (concentration 80 mg/mL) was spin coated at 1300 rpm for 1 min, leading to the formation of a 500 nm thick dielectric layer, followed by an annealing at 80 °C for 10 minutes for residual solvent removal. To end the devices, a 30 nm thick aluminium layer was deposited on the channel area by thermal evaporation. Devices were then annealed at 130 °C in nitrogen atmosphere for 12 h.

**Electrical characterization.** Devices were characterized via a Keysight B1500A Semiconductor Parameter Analyzer in nitrogen atmosphere. For devices annealed at 150 °C, $\mu_{sat}$ was extracted at $V_{DS}$ = 60 V and $V_{GS}$ = 30 V, where the square root of the current in saturation regime is linear. For all the other devices $\mu_{sat}$ was extracted at $V_{DS}$ = $V_{GS}$ = 60 V.

**Supporting information**

Supporting information are available upon request to the authors.

**Acknowledgements**

R. M. and M. S. thank the DFG for funding (IRTG 1642 Soft matter science and project SO 1213/ 8-1). A. Warmbold and M. Hagios are acknowledged for measuring DSC and SEC. M. G., A. L. and M. C. acknowledges the support by the European Research Council (ERC) under the European Union's Horizon 2020 research and innovation program "HEROIC", Grant Agreement No. 638059.